\documentstyle[epsf,12pt]{article}
\begin{document}
\title{\vspace{-15mm}
       {\normalsize \hfill
       \begin{tabbing}
       \`\begin{tabular}{l}
  SLAC--PUB--6529 \\
	 PM/94-17  \\
	 June 1994 \\
  hep-ph/9406357 \\
	 (T)   \\
	\end{tabular}
       \end{tabbing} }
       \vspace{8mm}
\setcounter{footnote}{1}
 Fermion and Sfermion Effects\\
in $e^+ e^-$ charged Higgs pair production
\thanks{Work
supported in part by the Department of Energy,
contract DE--AC03--76SF00515.}}

\renewcommand{\thefootnote}{\fnsymbol{footnote}}
\vspace{5mm}

\author{
A. Arhrib, M. Capdequi Peyranere \\ {\normalsize \em
Laboratoire de Physique Math\'{e}matique}\\ {\normalsize \em
(Unit\'{e} Associ\'{e}e au CNRS n$^o$ 040768)}\\ {\normalsize \em
Universit\'{e} de Montpellier II Sciences et Techniques du Languedoc}\\
{\normalsize \em
Place E. Bataillon, Case 50 F-34095 Montpellier Cedex 5, France}\\[3mm]
\setcounter{footnote}{4} and\\ G. Moultaka
\thanks{On leave of absence from Laboratoire de Physique
Math\'ematique, France.
Work supported by the CNRS and by a grant from NSF.} \\
 {\normalsize \em
 Stanford Linear Accelerator Center} \\
{\normalsize \em Stanford University, Stanford, California 94309}}
\date{}

\maketitle

\begin{abstract}
Top and bottom one loop corrections to $e^+e^-\rightarrow H^+ H^-$
are considered at NLC energies and found to be significantly
large and negative over the full range of $tan \, \beta$.
Moreover, loop effects of moderately heavy squarks and sleptons in the
minimal\newline supersymmetric extension of the standard model, tend to
cancel partly or fully these large corrections, thus providing
a useful indirect information about the supersymmetric sector.
The overall effect can still range between $-25 \%$ and $+25 \%$,
however lies typically around $-10 \%$ in a wide
range of the model parameters.
\end{abstract}

\vfill
\begin{center}
Submitted to {\em Physics Letters B}
\end{center}
\vfill
\newpage
{\bf Introduction.}\\[.15in]
 The scalar sector of the Standard Model (SM)
 still lacks direct experimental evidence.
 Yet the recent LEP precision measurements, together with the dogma
of grand unification seem to favour not just an extended Higgs sector,
but rather a full-fledged supersymmetric extension of the SM [1].

Furthermore, it appears,[2] , that
the minimal supersymmetric extension of the SM, MSSM [3],
can be at least as successful as the SM itself,
when confronted with the LEP data, even for not too heavy
susy particles.
All this, together with the general strong
theoretical motivation for the MSSM when
viewed as the low energy effective
theory of some grand unified supergravity theory [4], makes it rather
compelling to expect that the next generation
of $e^+e^-$ machines (LEPII, NLC) and/or
hadron colliders (LHC and possibly an upgraded Tevatron)
could discover neutral and/or charged
Higgs scalars [5,6], and perhaps establish first evidence
for the existence of SUSY particles.\par
However even then, one might still have to face the situation
where a substantial part of the SUSY spectrum remains unidentified.
Typically, squarks and gluinos
can in principle be discovered at future
hadron colliders up to masses in the
$1-2$Tev range while sleptons would
become invisible to LHC if heavier
than $\sim 250$GeV  or so [7], due to their weak coupling
and a prominent background. It is
then natural to ask to what extent can the
phenomenological study of the Higgs sector
give complementary, though indirect,
information about susy spectrum. Furthermore it is clearly
important to investigate the extent to which loop corrections can
modify the tree-level based assessment of expected Higgs production
rates at future machines.
\par
Hereafter we study this question in
the context of charged Higgs production at
the projected $e^+e^-$ machine with c.m.
energy $\sqrt s = 500GeV$.\par
The present experimental lower bounds on $m_{H^{\pm}}$ lie around
$35-40$ GeV [8], (to be contrasted with those of the Standard
Higgs, 60 GeV [9], and the neutral MSSM Higgses, 44 (resp. 20) GeV
for h (resp. A) )[10]. Recently there has also been some renewed
interest in constraints on $m_{H^{\pm}}$ which can be inferred from
the present upper bound on the branching ratio
$Br(b\rightarrow s \gamma)$ [11], where $H^\pm$ as well as
charginos and stops enter at the one-loop level, [12].
However a charged Higgs lighter than 250 GeV seems to be still
allowed in the context of the MSSM, [13], even though the region
under 200GeV is rather not favoured in a general analysis of susy
unification constraints
\footnote{We thank L. Roszkowski
for a discussion on this point} [14].
If so, one would still expect typically a few 100 events from the
pair production at NLC energies, assuming a nominal luminosity
of 10 $fb^{-1}$ and $M_{H^\pm} \sim 200 - 230$ GeV, [15].
It is worth recalling here that
the charged Higgs production
via the $Z^o$ or the photon at the tree level
has the peculiar feature of being independent of the non-standard
parameters
($m_A$, tan$\beta$, $m_{\tilde{f}}$, $m_{\tilde{q}}$, $\cdots$)
except of course for the charged Higgs mass itself.
(One should keep in mind the $\tan \,\beta$ dependence at tree level
in the charged Higgs leptonic and hadronic partial decay widths.
In the present study however, we concentrate exclusively
on the production process).
In the process $e^+e^- \rightarrow H^+H^-$ all non-standard
parameters will first occur at the one loop level and the hope is
that this would increase the sensitivity to the susy spectrum
itself, once the well known SM one loop corrections are properly
subtracted.
This would complement the information from the
neutral Higgs sector where the free parameters at the tree
level render the analysis somewhat more intricate [16].
 It thus appears
that quantum corrections to $e^+e^- \rightarrow H^+H^-$ could
provide useful indirect indications for a heavy susy spectrum.
However one still has to control the potentially large top-bottom
contributions which are sensitive to $tan \, \beta$, the relative
strength of the Higgs vacuum expectation values in two Higgs doublets
models.\par
In this letter we concentrate on the estimate of the effects
from the standard heavy fermions and their susy scalar partners which
occur in self-energies and vertices on the integrated cross-section,
leaving the inclusion of the full radiative corrections
from the MSSM to
a forthcoming paper. The leading (top-bottom) fermionic contributions
will depend solely on $\tan \, \beta$ as a non-standard free parameter
and are actually those of a (type II) two-Higgs doublets model
without supersymmetry. Those of the scalar partners bring in the
dependence on the soft susy breaking parameters, the left-right
mixing angles, the squark and slepton masses, etc...
As it will turn out, the effect from top-bottom will be
at least a $-8 \%$ to $-10 \%$ dip in the cross section,
for $m_{top} \sim 165-175GeV$ (but actually much larger
if $\tan \,\beta <5$ or $\tan \, \beta > 10$),
 while that from squarks and sleptons is found to be generically positive.
However the possibility of large cancellation between  the above
contributions is possible only in a marginal region of parameter space.
A generic $-10 \%$ overall effect is found over a wide range of the
parameter space and for moderately heavy squarks and sleptons.
Yet there will still be regions where
the effect is much larger and with either signs.\par

Before discussing the results in more detail, we describe briefly
in the next section the renormalization scheme and the
parametrization we use.\\[.15in]

{\bf Renormalization and parametrization}\\
At tree level, the leading contribution (i.e. neglecting the electron
mass)
to the angular distribution corresponds
to the Feynman diagram  drawn below, and is given by,
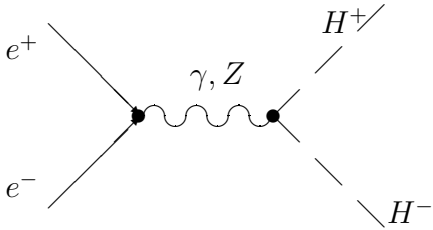
\def\fe#1#2{
\begin{picture}(100,200)(0,0)
\put(34,0){\circle*{5}}
\put(0,35){\vector(1,-1){35}}
\put(-10,26){\makebox(0,0){#1}}
\put(-10,-26){\makebox(0,0){#2}}
\put(0,-35){\vector(1,1){35}}
\end{picture} }
%
\def\bp#1#2{
\begin{picture}(100,200)(0,0)
\multiput(0,0)(16,0){3}{\oval(8,8)[t]}
\multiput(8,0)(16,0){3}{\oval(8,8)[b]}
\put(#2,15){\makebox(0,0){#1}}
\end{picture}   }
\def\fsp#1#2{
\begin{picture}(100,200)(0,0)
\put(0,0){\circle*{5}}
\multiput(0,0)(16,16){3}{\line(1,1){10}}
\multiput(0,0)(16,-16){3}{\line(1,-1){10}}
\put(27,36){\makebox(0,0){#1}}
\put(52,-36){\makebox(0,0){#2}}
\end{picture}   }
\begin{center}
\begin{picture}(100,100)(0,0)
\put(0,50) {\fe{$e^+$}{$e^-$}}
\put(40,50) {\bp{$\gamma,Z$}{24}}
\put(85,50) {\fsp{$H^+$}{$H^-$} }
\end{picture}
\end{center}
\begin{equation}
\left(\frac{d \sigma}{d \Omega}\right)_0=\frac{\kappa}{64 \pi^2
s}\left(\sum_{spin}|\frac{1}{2}{\cal M}_{Born}|^2\right) ,
\end{equation}
with
\begin{equation}
\sum_{spin}|\frac{1}{2}{\cal M}_{Born}|^2=
\frac{1}{2}e^4 \kappa^2 \left(1+g_H^2
\frac{g_V^2+g_A^2}{(1-m_Z^2/s)^2}-\frac{2 g_H g_V}{1 -m_Z^2/s}\right)
sin^2 \theta ,
\end{equation}
$\kappa ^2=1-\frac{\textstyle 4 m_{H^\pm}^2}{\textstyle s}$, $m_{H^\pm}$
 the mass of the charged Higgs, $m_{Z}$
 the mass of the neutral Z boson, s the
energy in the center of mass frame
, $g_V=(1-4 s_w^2)/(4 c_w s_w)$, $g_A=1/(4 c_w s_w)$,
$g_H=-(c_w^2-s_w^2)/(2 c_w s_w)$, $c_w\equiv\cos\,\theta_w$,
$s_w\equiv sin\, \theta_w$
and $\theta$ the scattering angle.\\
The total cross-section is, [15]:
\begin{equation}
\sigma_0=\frac{\pi\alpha^2\kappa^3}{3 s}
\left(1+g_H^2 \frac{g_V^2+g_A^2}
{(1-m_Z^2/s)^2}-\frac{2 g_H g_V}{1-m_Z^2/s}\right) .
\end{equation}
\par The parameters entering the above tree level observables are
all standard except for the charged Higgs mass.
Furthermore, the non-standard parameters which will appear at
the one--loop level can be consistently taken as bare in our computation.
In particular issues like the renormalization scheme dependence
of the separation between large and small $\tan \, \beta$ which occurs
in the case of neutral Higgs [17] can be ignored in our case as being
a higher ordre effect.\\
On the other hand, in the process of comparing the non-susy case,
(i.e. mainly the top-bottom contributions) with the MSSM case,
(i.e. fermions plus their scalar partners), one should in principle
take into account the fact that in the first case all Higgs masses
are independent quantities while Higgs mass sum rules exist in the
second case.\\

Let us recall here that these tree-level relations [18], can
be largerly violated due to radiative corrections, in the case
of neutral Higgses [19], whereas the relation involving the charged
and neutral CP-odd Higgses was shown to be much less sensitive
to loop corrections, apart from a small region in parameter space [20].\\

Because of such Higgs mass relations one is not at liberty to choose
the Higgs mass counterterms such that {\it all} the physical Higgs masses
be identified with the tree-level corresponding parameters in the
lagrangian. For instance in ref.[17] the above choice is made for the
neutral CP--odd Higgs. The renormalized masses of the other Higgses
are then automatically determined by the corresponding sum rules without
further subtraction conditions, (provided of course the standard Z and W
masses have been renormalized). In the present paper we also adopt an
on-shell scheme as defined in ref.[22], however we choose to identify
the physical charged Higgs mass with the corresponding parameter in
lagrangian, i.e., the mass counterterm is given by
\begin{equation}
\delta m^2_{H\pm} = Re \, \sum_{H^+H^+} (m^2_{H^\pm}) \end{equation}
where $\sum_{H^+H^-} (p^2)$ is the charged Higgs bare self-energy.
The above choice is the simplest in our case since we consider
only the charged Higgs and do not look at constraints from the
violation of the tree level relation between $m_{H^\pm}$ and $m_A$.
Indeed the only sizeable constraint seems to come from the
light charged Higgs scenario
($m_{H^\pm} < M_W$) [20], which we disregard here.
On the other hand eq.(4) corresponds to the
natural on-shell prescription in the non susy case and in any case
implies that the one-loop correction to the $H^\pm$ self-energy
cancels out when the Higgs pair is produced on shell. All the remaining
divergencies are then absorbed in the renormalization of
the electric charge,
$M_Z, M_W, m_e$ and the wave functions following [22].
Hence the renormalization procedure will involve essentially
a set of (standard) parameters, which would facilitate a simultaneous
treatment and comparison of susy and non-susy cases.
The purely standard contributions to these
corrections can be obtained by running their known
expressions up to NLC energies. ( The treatment of the
hadronic contribution to the self-energies would require
however the knowledge of $e^+e^- \rightarrow \gamma^*
\rightarrow hadrons$ up to these energies).
These purely standard effects will not be included in the
present study, except for the non-universal contributions
coming from the heavy top-bottom doublet.
There are also the pure susy contributions to the standard
parameters. In the energy range $500 GeV-1TeV$, one no more
expects all the susy particles to decouple, since some or all of the
charginos and neutralinos are expected to be much lighter.
The loop contributions of charginos and neutralinos to the
bosonic self-energies and the $(\gamma)ZH^+H^-$ vertices
can be comparable to those from other sectors.
(Note however that in the present case the potential
enhancement of their coupling to the lepton-slepton sector for
large $tan \, \beta$ is suppressed by the electron mass).
Only the squark and slepton corrections enjoy
further enhancement from the free parameters $\mu, A_f, tan\,\beta$,
which are the ones we are interested in here.
In this case the $(\gamma)ZH^+H^-$ vertex counterterms are related
to the Higgs wave function renormalization (see also [23]).

In the more general case however, the vertex counterterms will be also
related to the one loop renormalization
of the photon and Z self energies.
\footnote{ as for the infrared, part it will be included in the full
one-loop computation elsewhere, [21].}.
Accordingly, the full vertex counterterm is given by,
(using notations of ref. [22])
\begin{eqnarray}
& &\delta (A_\mu H^+ H^-)=-i e [\delta Z^{H^{\pm}}+
(\delta Z_1^{\gamma}-\delta Z_2^{\gamma})
+g_H (\delta Z_1^{\gamma z}-\delta Z_2^{\gamma z}) ]
(p+q)_\mu\nonumber\\
& &\delta (Z_\mu H^+ H^-)=i e g_H [\delta Z^{H^{\pm}}+
(\delta Z_1^{z}-\delta Z_2^{z})
+\frac{1}{g_H}(\delta Z_1^{\gamma z}-\delta Z_2^{\gamma z}) ]
(p+q)_\mu
\end{eqnarray}
\noindent
 all other MSSM parameters being consistently kept at their tree level
values.\\
Specifying to the quarks, leptons and their scalar partners, the relevant
one-loop diagrams are depicted in fig.8 and
have been generated and computed
as part of the full one loop corrections in the 't Hooft-Feynman
gauge, using FeynArts and FeynCalc packages [24] supplied with a
full MSSM Feynman rules code [25]. We also used the Fortran FF-package
 [26], in the numerical analysis.\par
It is worth noting that the set of diagrams in fig.8 is gauge invariant
by itself and thus can be studied separately.
(Note also that the 1PI contribution to $Z(\gamma)H^+H^-$ with
$\tilde{q} \tilde{q} H^+ H^-$ coupling is trivially vanishing).
Apart from the top quark mass, we have
{\it generically} nine non-standard
free parameters to deal with, which one can choose as,
\begin{itemize}
\item the charged Higgs mass $m_{H^{\pm}}$.
\item the ratio of the
two vacuum expectation values tan$\beta$.
\item  the $\mu$-parameter parametrizing the (susy invariant)
Peccei-Quinn symmetry breaking term in the superpotential.
\item the squark and slepton masses $m_{\tilde{q}_{1,2}}$,
$m_{\tilde{l}_{1,2}}$.
\item the left-right mixing angles $ \theta_{u,d} $ and $\theta_l $.
\end{itemize}
Furthermore we will neglect CKM and super-CKM flavor mixing effects.
The soft supersymmetry breaking parameters $A_f$ are then connected to
the previous ones through,
\begin{equation}
A_{(u,d)} =\frac{\sin\,( \theta_{(u,d)}) \,\, \cos\, (\theta_{(u,d
)})}{m_{(u,d)}}
( m_{(\tilde{u}_1,\tilde{d}_1)
}^2 - m_{(\tilde{u}_2, \tilde{d}_2)}^2 ) -
\mu \, (cot\beta,tan\beta )
\end{equation}
where $u$ and $d$ denote up and down flavors for quarks or
leptons and $m_{(u,d)}$, $m_{(\tilde{u}_1,\tilde{d}_1)}$ and
$m_{(\tilde{u}_2,\tilde{d}_2)}$
are the {\sl physical} masses of the respective scalar partners
.(In eq.(5) we absorbed the soft susy breaking mass parameter
``$m_6$'' in the definition of $A_{(u,d)}$, see ref.[3] for notations ).
On the other hand we will carry later discussions in terms of
degenerate or non degenerate scalar partners
(at least for the scalar tops), rather than in terms of squark
(or slepton) mixing. Among others, a significant consistency
check of our one--loop expressions is their independence of
$\theta _u$ (resp. $\theta_d$) when the up--squarks (resp.
down--squarks) of a given family are mass degenerate.

The proliferation of free parameters is of course due to the fact that
we do not assume any specific underlying theory at higher energies.
However in the numerical study we will generally remain close to
the typical range of values suggested by the underlying assumptions
of supersymmetric grand unification
and radiative electroweak breaking[28].\\
For completeness let us simply recall how the free parameters
enter the charged Higgs couplings to quarks, leptons
and their scalar partners, in the left-right basis,
 (see ref.[3,18] for more details) :
\begin{equation}
H^{\pm} u d : \,\, \frac{i g}{2 \sqrt{2}m_W}( m_d tan\beta + m_u cot\beta
\pm ( m_d tan\beta -  m_u cot\beta) \gamma_5 )
\end{equation}
\begin{equation}
H^+ \tilde{u}_{(R,L)} \tilde{d}_{(L,R)} :\,\, H_{(RL, LR)} \equiv
\frac{-i g}{\sqrt{2}}\frac{m_{(u,d)}}{m_W}
(\mu - (A_u cot\beta, A_d tan\beta) )
\end{equation}
\begin{equation}
H^+ \tilde{u}_L \tilde{d}_L : \,\, H_{LL} \equiv \frac{-i g}{\sqrt{2}}
m_W (sin(2\beta) - \frac{m_d^2 tan\beta + m_u^2 cot\beta}{m_W^2} )
\end{equation}
\begin{equation}
H^+ \tilde{u}_R \tilde{d}_R : \,\, H_{RR} \equiv \frac{-i g}{\sqrt{2}}
\frac{m_u m_d}{m_W} (tan\beta + cot\beta)
\end{equation}
Furthermore the left-right mixing angles, $\theta_q$, $\theta_l$
will enter the game  whenever $\tilde{q}_L (\tilde{l}_L)$ and
$\tilde{q}_R (\tilde{l}_R)$ are not mass eigenstates.\par
{}From the above dependence on the various free parameters
one can understand the qualitative features of the results
discussed in the next section. It is however already clear
for instance that in general the bottom quark mass should
not be neglected in the $H^{\pm} t b$ vertex, since its
contribution becomes comparable to that of the top
($m_t \simeq 160$GeV) as soon as tan$\beta \geq 6$.
On the other hand the chirality non-conserving $H\tilde{u} \tilde{d}$
couplings become insensitive to the scalar partner masses in the
degenerate case, but remain very sensitive to the $\mu$
parameter. In contrast the chirality conserving ones
are controlled by the fermion masses and $tan(\beta)$ only.

{\bf Results and discussion}\\
Since we are interested here mainly in the trend of the corrections
as a function of the free parameters, we will consider only
the leading effect from the interference between the one loop and
tree level contributions. We do not include any
improved resummation and the standard model parameters will be kept
at their tree level values rather than being run to the energy
scale under consideration. This approximation is safe inasmuch as
it somewhat underestimates the effect from the running of $\alpha_{QED}$.
\\[.15in]
Perhaps more importantly, and since we are dealing with very large
effects in some cases, it is important to have some criteria
which allow to control the validity of the perturbative result.
With this respect a more elaborate treatment using
renormalization group techniques to include higher order leading
logs would be appropriate. This might extend the region in
parameter space, overwhich the perturbative results remain reliable.
In the present study, however, we will adopt the following simple
(and preliminary) criterion.
We require that the contributions to $\tilde{\delta}_t$,
(see eq.(11) below), which are $O(\alpha^2)$ and obtained
from squaring the one-loop amplitude, do
not exceed $50 \%$ of the $O(\alpha)$ ones.
Although incomplete, this criterion does give some constraints
as we will see, on the otherwise increasingly big effects in some
regions of the parameter space. It will constrain as well the
range of variation of the free parameters, over which we consider
the (unimproved) one-loop results to remain meaningful.
For instance allowing $tan \, \beta$ to be in the range
$1< tan \, \beta < m_t/m_b$ [27], our criterion will most of the time
constrain further the valid domain as we will see. The same
holds for $\mu$ which we will take a priori between 0 and 500GeV.
\\[.15in]
On the other hand the effects we are interested in
appear in the integrated
cross-section. There are no shape effects in the angular distribution
since the self-energy and vertex corrections clearly contribute
only to the leading tree-level s-wave component with the typical
$\sin^2\, \theta$ dependence
\footnote{angular effects arise only from box contributions
which will be considered in [21].}.
The one-loop correction
$ \tilde{\delta}_t $ is
defined through
\begin{equation} \sigma = \sigma_0\,(1 + \tilde{\delta}_t)
\end{equation}
where $\sigma_0$ is the tree-level cross-section given by eq.(3)
and
$$\tilde{\delta}_t \equiv \tilde{\delta}_t(m_t, tan\,\beta, m_{H^{\pm}}
, \mu, m_{\tilde{q}_{1,2}}, m_{\tilde{l}_{1,2}}, \theta_{\tilde{q}},
\theta_{\tilde{l}} )$$
As we will see, qualitatively significant information can be obtained
in spite of the large number of free parameters especially as regards
the potentially large effects that may arise for large $tan\, \beta$
and/or soft SUSY breaking scale. In fig.1 we show the expected
tree level integrated cross section $\sigma_0$ as a function
of $m_{H^{\pm}}$. Most of the subsequent discussion of the radiative
corrections will be carried with $m_{H^{\pm}}= 220$ GeV, a value which
lies in the typical range
compatible with the indirect constraints from
$BR(b \rightarrow s \gamma)$ and in the same time leading
to an interesting production rate at a $500$GeV ($10 fb^{-1}$) machine.
\\
The contributions directly sensitive to $tan \, \beta$ and to the soft
susy breaking parameters in $\tilde{\delta}_t$ come
from matter fields and their susy partners in the $\gamma H^+ H^-$
and $Z H^+ H^-$ vertices. We also included the contributions
to the $\gamma$ and $Z$ self-energies from top, bottom and
squarks and sleptons loops, but these remain actually numerically
small. (As we mentioned before, we do not include in the
present study all the remaining one loop corrections to
$e^+e^- \rightarrow H^+ H^-$ as they are either purely standard
and in principle under control, or correspond to those susy
contributions which cannot be enhanced by large free parameters.
The same turned out to be true for the potentially large contributions
from the various neutral and charged Higgses. These do not exceed
a few percent as long as the Higgses masses remain far below the
TeV scale). \\
The leading $O(\alpha)$ correction can be written in the form
\begin{equation}
\tilde{\delta}_t \equiv
\frac{\alpha \, \kappa^3}{12 \sigma_0}
Re\{ (-1 + \frac{g_H g_V}{1-m_Z^2/s})(\frac{1}{s} X_{\gamma}
+\frac{(-1 + 4 s_w^2)}{s - m_Z^2} X_Z) - \frac{g_H g_A}{1-m_Z^2/s}
\frac{X_Z}{s- m_Z^2} \} \end{equation}
where $X_{\gamma}$ (resp. $X_Z$) collect the one-loop corrections
to the $\gamma H^+ H^-$ (resp. $Z H^+ H^-$) renormalized vertices
as well as the $\gamma$ and $Z$ self-energies and we drop the
$Z$ width in the propagator.
(We also absorb
in $X_\gamma$ and $X_Z$ the corresponding couplings to the initial state).
The full analytic expressions of the leading contributions, namely
the renormalized vertex corrections due to top-bottom and stop-sbottom
loops are given in the appendix.
The leading effect from the top and bottom quarks is shown in fig.2a
and 2b. as a function of $m_t$ or $tan\, \beta$ (with $m_b \simeq
4.5GeV$). The contribution turns out to be always negative
and significant, in the full $tan \, \beta$ range. It reaches at
least $-8\%$ to $-10 \%$ if $m_t \geq 165$ GeV for $tan \, \beta$
between $5$ and $10$
and becomes much larger away from this range. Also the fact that
 $\tilde{\delta}_t$ is bounded from above (for a given value of
$m_t$) is a direct consequence of the form of the $H^{\pm} t b$
coupling in eq.(6), and corresponds to the value of $tan \, \beta$
where the top and bottom effects in $H^{\pm} t b$ vertex become
comparable. In any case the main feature here is that the top-bottom
effects are generally significantly large and {\sl negative} even
for not too large $tan \, \beta$ and that in a non-susy two
Higgs doublet model, they can easily overwhelm the remaining one
loop corrections and lead to a significant dip in the expected
charged Higgs production rate. Of course the effect can become
eventually too large ($\sim -50 \%$)
for very large $tan \, \beta$ so that one
should start worrying about higher order corrections.
Actually, according to our perturbation criterion, $O(\alpha^2)$
effects can become sizeable if $tan \, \beta > 20$ or
$tan \, \beta <1.5$
in which case $\tilde{\delta}_t$ reachs respectively $-25\%$ and
and $-20 \%$.
This suggests that
(conservatively) the magnitude of the real effect would not exceed
the values quoted above, leading anyway to a dip in the total
cross-section.
For a further assessment of the effect we show in
fig.3 the sensitivity to the charged Higgs mass. \par

The situation becomes drastically different in the susy case.
As illustrated in fig.4a-b, the contributions of squarks and sleptons
are generically positive and become increasingly significant
for large $\mu$ and $tan \, \beta$ as far as squark and slepton masses
are not in the TeV range. Note that we assumed all squarks (resp.
sleptons) degenerate in mass, which is a typical feature when
running these masses  down from a common value at a higher unification
scale [28], except for the susy partners of the top quark which are
non-degenerate. We will consider this more realistic case later on.
For now it is important to stress that the squark and slepton
contributions can largely reduce the top-bottom effect. This
is illustrated in fig. 5 for various values of $\mu$ and $tan \, \beta$.
(Note that since squarks (resp. sleptons) are taken degenerate in mass,
the parameter $A_f$ in eq.(5) is no longer free, also the physical
observables do not depend in this case on the actual value of the
(arbitrary) left-right mixing angles $\theta_q$ and $\theta_l$).\\
As one can see from fig.5 , the inclusion of squarks and sleptons,
(actually mainly the susy partners of $(t,b)$ and $(\nu, \tau)$)
levels the overall negative effect to about $-10 \%$, for a wide
range of $tan \, \beta$, but a complete cancellation is also
possible and even a significant overall positive effect. The details
depend of course on the chosen values of $m_{\tilde{l}}, m_{\tilde{q}},
m_t$ and $m_{H^\pm}$, however varying $\mu$ and $tan \, \beta$
simultaneously and assuming $m_{\tilde{l}}$ and $m_{\tilde{q}} > 250$GeV
one finds a large and negative effect
over a fairly large region of parameter
space. This is illustrated by the contour plots of fig.9 where we took
for illustration $m_{\tilde{l}} = 300$ GeV and $m_{\tilde{q}} = 400$GeV.
In this case a negative effect reaching down to $-25 \%$ is expected
in a large domain comprising virtually any value of $\tan \, \beta$ and
$\mu$, although the larger $\tan \, \beta$ the smaller $\mu$ should be.
In contrast, positive overall effects, reaching up to $+25 \%$
are squeezed in a much smaller region corresponding to $\tan \, \beta$
and $\mu$ simultaneously large. Note also that a cancellation between
the top--bottom and squark-slepton contributions occurs only in a
pencil-like region of large parameters.
(The maximum values $\pm 25 \%$ correspond to the maximum effect allowed
within our perturbativity criterion which requires in this case
$\tan \, \beta < 25$ ( resp. $<30$) ifi $\mu < 50 $GeV (resp. $>400$GeV).
Furthermore the region between say, $-8 \%$ and $-12 \%$ tends to be
much less sensitive to the values of $\mu$ and $\tan \, \beta$ than the
rest of the negative region, while for $\tan \,\beta < 5$ the leading
effect is from top--bottom loops.
We turn now briefly to two other possibilities, namely,
a) squarks are very heavy ($\sim 1$ TeV) and sleptons much lighter
but heavier than $\sim 250GeV$; b) the susy partners of the top
are non-degenerate ($ m_{\tilde{t_1}} - m_{\tilde{t_2}}
\sim 100 -300 GeV$);
both cases a) and b) would modify the previous discussion.\\ As we
mentioned previously, case a) corresponds to a situation where the
squarks can still be detected at an LHC machine but not the sleptons.
In addition in this case the squarks contibution to $\tilde{\delta_t}$
becomes negligible (see fig. 4a), a direct consequence of the decoupling
of the heavy susy sector from ``low energy'' physics [29], as far as
the parameter $\mu$ present in the Higgs-squark-squark vertex is not
very large ($\mu <500$ GeV). In such a scenario a slepton in the
$300-400$GeV mass range could still give a significant effect. For
instance as it can be seen from fig.6 and fig.2b, a 300 GeV scalar-tau
can reduce the (negative) top-bottom effect down to $-15 \%$
if $\mu \simeq 450$GeV and $tan \, \beta \simeq 30$. However taking into
account our criterion for perturbativity the acceptable effect would be
around only $4 \%$.
Although this might
be a marginal possibility as far as the parameter space is concerned,
it remains worth considering as an indirect signature of sleptons heavier
than $250GeV$,
given the rather limited discovery potential at LHC, as well as
at a $500$ GeV $e^+e^-$ NLC machine \footnote{ actually even a
laser back scattered photon could scan only selectrons
somewhat above the beam energy, through
$e \gamma \rightarrow \tilde{e} \tilde{\gamma}$, but not $s-taus$
which give the leading effect in our case.}.\par

Finally let us comment briefly case b).
Note that in this case the soft susy breaking parameter $A_t$
(or equivalently the left-right mixing angle $\theta_t$ between
the two scalar partners of the top quark, see eq.(5)) becomes
a free parameter.The effects are found to remain comparable
to those in the degenerate case as long as
$\theta_t$ is small ($ <0.2$rad.),
but they are generally very sensitive to this angle.
In fig.7a--b we illustrate the squarks contribution for two values
of the stop mass splitting with $\theta_t=0.4$ rad. and where we
assumed for definiteness the lighter stop mass to be above $150$ GeV
and $\mu = 250 $ GeV. (Note also that the sensitivity to the sign of
$\mu$ turns out to be weak for the chosen values in the figures).
It should be stressed that unlike the case of the one--loop
corrected mass relation between $m_{H^\pm}$ and $m_A$ [20], the effect
under study does not exhibit a new enhancing $m_t$ dependence in the
presence of left--right mixing. (see expressions in appendix b)).
In any case a full 3--parameter study becomes mandatory and is out
of the scope of this letter. However in the special case of figure 7a
for instance, one finds that the perturbative result remains reliable
even for $m_{\tilde{t_1}}$ as low as $300$GeV and $\tan \, \beta$
as large as $20$, in which case the total effect (including top-bottom
contribution) would be around $+ 30 \%$. Finally it is worth noting
that one can get even larger total positive effects, up to $50 \%$ which
remain rather safe from the perturbative point of view (the $O(\alpha^2)$
contributions being only a few percent of the $O(\alpha)$ ones) typically
if the left--right mixing angle is $\sim \pi/2$ and $\tan \, \beta
\sim 30$.\par
To conclude, we showed in this letter that potentially large and model--
dependent one--loop corrections to the (model--independent) tree--level
charged Higgs production cross section can occur,
for generic values of the model parameters,
at $\sqrt{s}=500$ GeV $e^+e^-$ collisions. These should be taken into
account when discussing production rates at NLC.
A further investigation
at higher NLC energies and for heavier charged Higgses is however needed,
as well as the inclusion of loop effects from the rest of the particle
spectrum and eventually an improved resummation in the region where
the effects are very large.

\newpage
\begin{center}
{\bf References}
\end{center}
\begin{tabbing}
\=qqqq\= \kill \\
\> [1] \>  U.Amaldi, W. de Boer and H.F\"urstenau, Phys.Lett. B260 (1991)
447;\\
\>     \>  J.Ellis, S.Kelley and D.V.Nanopoulos, Phys.Lett. B249 (1990) 441;\\
\>     \>  P.Langacker and N.Polonsky, Phys.Rev. D47 (1993) 4028;\\
\> [2] \>  G.Altarelli, R.Barbieri and F.Caravaglios, Phys.Lett. B314 (1993)
357;\\
\> [3] \>  H.E.Haber and G.L.Kane, Phys.Rep. 117 (1985) 75; \\
\>     \>  J.F.Gunion and H.E.Haber, Nucl.Phys. B272 (1986) 1;\\
\> [4] \>  For a recent review see "Supersymmetry and Supergravity :\\
\>     \>  Phenomenology and grand Unification'', R.Arnowitt and P.Nath,\\
\>     \>  SSCL-Preprint-503, and references therein;\\
\> [5] \>  ECFA Large Hadron Collider Workshop, Eds. G.Jarlskog and D.Rein,\\
\>     \>  CERN 90-10, vol.II page 605;\\
\>     \>  A.Stange, W.Marciano and S.Willenbrock, Phys.Rev.D49 (1994) 1354,\\
\>     \>  and preprint ILL-TH 94-8;\\
\> [6] \>  Proceedings of the Workshop on $e^+e^-$ Collisions at $50$ GeV :\\
\>     \>  The Physics Potential, ed. P.M.Zerwas, DESY 92-123;\\
\> [7] \>  F. del Aguila and Ll.Ametller, Phys.Lett. B261, 3, (1991) 326;\\
\> [8] \>  ALEPH collab., Phys.Lett.B241 (1990) 623;\\
\>     \>  DELPHI collab., Phys.Lett.B241 (1990) 449 ;\\
\>     \>  L3 collab., Phys.Lett.B252 (1990) 511;\\
\>     \>  OPAL collab., Phys.Lett.B242 (1990) 299;\\
\>     \>  UA2 collab., Phys.Lett. B280 (1992) 137;\\\
\> [9] \>  ALEPH collab., Phys.Lett.B313 (1993) 299;\\
\> [10] \> ALEPH collab., Phys.Lett.B313 (1993) 312;\\
\> [11] \> CLEO collab., Phys.Rev.Lett. 71 (1994) 674;\\
\> [12] \> S.Bertolini et al., Nulc. Phys. B353 (1991) 591;\\
\>      \> R.Barbieri and G.Giudici, Phys.Lett.B309 (1993) 86;\\
\>      \> J.L.Hewett, Phys.Rev.Lett. 70 (1993) 1045; V.Barger, M.S.Berger \\
\>      \> and R.J.N.Phillips, Phys.Rev.Lett. 70 (1993) 1368;\\
\> [13] \> R.Garisto and J.N.Ng, Phys.Lett. B315 (1993) 372;\\
\> [14] \> G.L.Kane et al., UM-TM-93-24;\\
\> [15] \> S. Komamiya, Phys.Rev. D38 (1988) 2158;\\
\> [16] \> A.Brignole et al.in ref. [6] page 613;\\
\>      \> A.Djouadi, J.Kalinowski, P.M.Zerwas, Z.Phys.C57 (1993) 569;\\
\>      \> and in ref.[6] page 83;\\
\> [17] \> P.H.Chankowski, S.Pokorski and J.Rosiek, Phys.Lett.B286 (1992)
307;\\
\>      \> and preprint MPI-PH-92-117, hep-ph/9303309 (unpublished);\\
\> [18] \> For a review see e.g. J.F. Gunion, H.E. Haber, G.L. Kane and S.
Dawson,\\
\>     \>  The Higgs hunter's guide (Addison-Wesley, Redwood City, 1990).\\
\> [19] \> H.E.Haber and R.Hempfling, Phys. Rev. Lett. 66 (1991) 1815; \\
\>      \> Y.Okada, M.Yamaguchi and T.Yanagida, Prog. Theor. Phys. Lett. 85
(1991) 1;\\
\>      \> J.Ellis, G.Ridolfi and F.Zwirner, Phys.Lett. B257 (1991) 83;\\
\>      \> for further references see the talk by F.Zwirner, CERN-TH.6792/93;\\
\> [20] \> A.Brignole et al., Phys.Lett. B271 (1991) 123; B273 (1991) 550(E);\\
\>      \> M.A.Diaz and H.E.Haber, Phys.Rev.D45 (1992) 4246;\\
\>      \> P.H.Chankowski, S.Pokorski and J.Rosiek, Phys.Lett.B274 (1992)
191;\\
\>      \> A.Brignole, Phys.Lett. B277 (1992) 313;\\
\> [21] \>  A. Arhrib, M. Capdequi Peyranere and G. Moultaka, in preparation.\\
\> [22] \> M.B\"ohm, W.Hollik and H.Spiesberger, Fortschr.Phys.34 (1986) 11;\\
\> [23] \> R.Foot, H.Lew and G.C.Joshi, Z.Phys. C47 (1990) 269;\\
\> [24] \> H.Eck and J. Kublbeck, Guide to FeynArts 1.0, University of
Wurzburg,
           1992.\\
\>      \> R.Mertig, Guide to FeynCalc 1.0, University of Wurzburg, 1992.\\
\> [25] \> A.Arhrib, thesis (to appear);\\
\> [26] \> G.J. van Oldenborgh, Comput.Phys.Commun. 66 (1991) 1;\\
\> [27] \> G.F. Giudici and G.Ridolfi, Z. Phys. C41, (1988) 447;\\
\>      \> M.Olechowski and S.Pokorski, Phys.Lett. B214 (1988) 393;\\
\> [28] \> See for instance D.J.Castano, E.J. Piard and P. Ramond,\\
\>      \> Phys.Rev.D49 (1994) 4882;\\
\> [29] \> J.F. Gunion and A. Turski , Phys.Rev.D 40, (1989), 2325; \\
\end{tabbing}
\newpage
\begin{center}
Appendix
\end{center}
a) top-bottom contributions to the $\gamma H^+H^-$ and
$ZH^+H^-$ vertices :\\
Using the following intermediate functions,
$$ f(x,y,z)= s+ 2 (x+y-z)$$
$$ g(x,y) = x^2 - y s - (m^2_H - y)^2$$
$$I(x) = 3(2 m^2_H - s) x - 6\,\cot^2\beta (\tan^2 \beta m^2_b + m^2_t)^2
$$
$$K_0(m^2_b, m^2_t)= (\tan^2\beta m^2_b+ \cot^2\beta m^2_t)
f(m^2_t,m^2_b,m^2_H) + 8 m^2_b m^2_t $$
$$K_1(x,y,m^2_1,m^2_2)= x (\tan^2\beta m^2_b + \cot^2 \beta m^2_t)
g(m^2_1, m^2_2) + y m^2_b m^2_t f(m^2_2, -m^2_1, m^2_H)$$
$$K_2^Z(t^2,m^2_1,m^2_2)=3 m^2_1 (m^2_2+m^2_H)
(\frac{m^2_1}{t^2} - t^2 m^2_2)
- 3 t^2 m^2_2(g(m^2_1,m^2_2)- m^4_1) - 3 \frac{m^6_1}{t^2} $$
$$K_2^{\gamma}(t^2, m^2_1, m^2_2) \equiv 0 $$
$$dB(m^2) = B_0(m^2_H,m^2_b,m^2_t) - B_0(s, m^2, m^2)$$
$$C(m^2_t, m^2_b) = C_0(m^2_H,m^2_H,s,m^2_t,m^2_b,m^2_t)$$
$$B_0'(m^2_H,...) \equiv \frac{\partial}{\partial \, s}
B_0(s,...) |_{s=m^2_H}$$
the (renormalized) vertex correction from the top-bottom loop
 can be written in the form, ($V=\gamma,Z$)
\begin{eqnarray}
\lefteqn{{\cal V}_V^{(top-bottom)}=}
\nonumber \\
& &\frac{N_c \delta^V \alpha^2}{c^2_w s^2_w M^2_Z
(s - 4 m^2_H)} [ (\delta_1^V K_0(m^2_b,m^2_t)+I(\cot^2\beta m^2_t))
dB(m^2_b)+ \nonumber \\
&~~~~~& \nonumber \\
& & (2 \delta_1^V K_0(m^2_b,m^2_t)+I(\tan^2\beta m^2_b))dB(m^2_t)+
\nonumber \\
&& \nonumber \\
& & 2 (K_1(x^V,y^V,m^2_b,m^2_t) + K_2^V(\cot^2 \beta,m^2_b,m^2_t))
C(m^2_t,m^2_b)+ \nonumber \\
&& \nonumber \\
\lefteqn{   4 (K_1(x^V,y^V,m^2_t,m^2_b)+ \frac{1}{2} K_2^V(\tan^2\beta,
m^2_t,m^2_b) - \frac{3}{2}m^2_b m^2_t f(m^2_b,-m^2_t,m^2_H))
C(m^2_b, m^2_t)+} \nonumber \\
&& \nonumber \\
& & 3 \delta^V_2 (s - 4 m^2_H)((\tan^2\beta m^2_b + \cot^2\beta m^2_t)
(m^2_H - m^2_b - m^2_t) - 4 m^2_b m^2_t) B_0'(m^2_H,m^2_b, m^2_t)]
 \nonumber \\
\end{eqnarray}
where $\delta^{\gamma} =\delta^{\gamma}_1=\delta^{\gamma}_2=1$,
 $\delta^Z=\frac{1}{8 c^2_w s^2_w}$,
$\delta^Z_1=x^Z= 2 s^2_w$,
$\delta^Z_2=(-1+ 2 s^2_w)$, $x^{\gamma}=1$, $y^{\gamma}=-2$ and
$y^Z= 3 - 4 s^2_w$; $N_c (= 3)$ is the color factor, $m_H, m_t$ and
$m_b$ denote respectively the charged higgs, the top quark and the
bottom quark masses. (The above expressions are valid for any up-down
quark doublet).
\\[.5cm]
b) Squark contributions to the $\gamma H^+H^-$ and $ZH^+H^-$ vertices :\\
In the following we assume the bottom squarks to be mass degenerate.
Using the following intermediate function,
\begin{eqnarray}
\lefteqn{{\cal F}^V_{\tilde{Q_i} \tilde{q_j}}=
 2 B_0(s, m^2_{\tilde{Q_i}}, m^2_{\tilde{Q_i}}) -
2 B_0(s, m^2_{\tilde{q_j}}, m^2_{\tilde{Q_i}}) } \nonumber \\
& &  + f(m^2_{\tilde{q_j}}, -m^2_{\tilde{Q_i}}, m^2_H)
    C_0(m^2_H,m^2_H,s,m^2_{\tilde{Q_i}},m^2_{\tilde{q_j}},
        m^2_{\tilde{Q_i}})   \nonumber \\
\end{eqnarray}
where $V= \gamma, Z$
we can write the (renormalized) vertex corrections from squark loops
associated to one quark generation (denoted generically by ``t" and ``b")
 in the following condensed form :
\begin{equation}
{\cal V}_V^{(squarks)} =
\frac{N_c \alpha^2 c_V}{s-4 m^2_H}( G^V_1
{\cal F}^V_{\tilde{t}_1 \tilde{b}}
                        + G^V_2{\cal F}^V_{\tilde{t}_2 \tilde{b}}
            - G^V_3{\cal F}^V_{\tilde{b} \tilde{t}_1}
  - G^V_4{\cal F}^V_{\tilde{b} \tilde{t}_2}
  + G^V_5 X + d_V (s- 4 m^2_H) \delta B)   \nonumber \\
\end{equation}
where $c_{\gamma}= 2$, $c_Z=1/(2 c_w s_w)$, $d_{\gamma}=1$,
$d_Z= (2 s_w^2 -1 )/(2 c_w s_w)$
and
\begin{eqnarray}
X=& 4 B_0(s,m^2_{\tilde{t}_1}, m^2_{\tilde{t}_2})
   -2 B_0(s,m^2_{\tilde{b}}, m^2_{\tilde{t}_1})
   -2 B_0(s,m^2_{\tilde{b}}, m^2_{\tilde{t}_2})& \nonumber \\
&+ f(m^2_{\tilde{b}}, -\frac{m^2_{\tilde{t}_1}+m^2_{\tilde{t}_2}}{2},
m^2_H)\, C_0(m^2_H,m^2_H,s,m^2_{\tilde{t_1}},m^2_{\tilde{b}},
        m^2_{\tilde{t_2}})& \nonumber \\
\end{eqnarray}
\begin{equation}
\delta B= h(\cos \, \theta_u, \sin \, \theta_u)
    B_0'(m^2_H,m^2_{\tilde{b}}, m^2_{\tilde{t}_1}) +
 h(\sin \, \theta_u, -\cos \, \theta_u)
    B_0'(m^2_H,m^2_{\tilde{b}}, m^2_{\tilde{t}_2})
\end{equation}
and also defined
\begin{equation}
G^V_1 =  a^V(-\cos^2 \,\theta_u, e_u)
 h(\cos \, \theta_u, \sin \, \theta_u)   \nonumber
\end{equation}
\begin{equation}
G^V_2 =  a^V(-\sin^2 \,\theta_u, e_u)
 h(\sin \, \theta_u,-\cos \, \theta_u)   \nonumber
\end{equation}
\begin{equation}
G^V_3 =  a^V(1, e_d)
 h_1(\cos \, \theta_u,\sin \, \theta_u)
 + a^V(0, e_d)
 h_2(\cos \, \theta_u,\sin \, \theta_u)   \nonumber
\end{equation}
\begin{equation}
G^V_4 =  a^V(1, e_d)
 h_1(\sin \, \theta_u,-\cos \, \theta_u)
 + a^V(0, e_d)
 h_2(\sin \, \theta_u,-\cos \, \theta_u)   \nonumber
\end{equation}
\begin{eqnarray}
G^Z_5 =  \frac{1}{2 c_w s_w} \cos \, \theta_u \sin \, \theta_u
[\cos \, \theta_u \sin \,\theta_u(H^2_{RL}+H^2_{RR}
-H^2_{LL} - H^2_{LR})&&\nonumber \\
 +(\cos^2 \, \theta_u - \sin^2 \, \theta_u)(H_{LL}H_{RL}+H_{RR}H_{LR})]
  \nonumber \\
\end{eqnarray}
where $e_u$ and $e_d$ denote the up and down quark electric charges
and $\theta_u$ the up left--right mixing and
$$G^{\gamma}_5 = 0 $$
$$a^{\gamma}(x,y)= y $$
$$a^Z (x,y) = \frac{x + 2 s^2_w y}{2 c_w s_w} $$
$$h_1(x,y) =  (x\, H_{LL} + y\, H_{RL})^2 $$
$$h_2(x,y) =  (x\, H_{LR} + y\, H_{RR})^2 $$
$$h(x,y) \equiv  h_1(x,y) + h_2(x,y) $$
$H_{LL}, H_{RL}, H_{RR}$ and $H_{LR}$ being the couplings
of the charged higgs to the various squarks (see definition in
the text).
Note in the above formulae the complete cancellation
of the $\theta_d$ angle due to the fact that the down squarks are mass
degenerate.
\newpage
\begin{center}
Figure Captions
\end{center}
Fig.1 : Integrated tree--level cross section as a function of the charged
Higgs mass.\\[.15in]
Fig.2 : top--bottom contribution (in percent) to the one--loop corrected
integrated cross section with $m_{H^\pm}= 220$ GeV, (a) as a function of
the top mass, for different values of $\tan \, beta$, (b) as a function
of $\tan \, \beta$ for three values of $m_t$.\\[.15in]
Fig.3 : One--loop top--bottom contribution (in percent) to the
integrated cross section as a function  of the charged Higgs mass,
for $m_t= 165$ GeV and three values of $tan \, \beta$. (the uppermost
curve corresponds to $\tan \, \beta = 10$.\\[.15in]
Fig.4 : (a) One--loop squarks contribution (in percent) to the integrated
cross section with $m_{H^\pm}= 220GeV$, $\mu= 250$ GeV and $m_t= 165$ GeV,
 as a function of the squark mass, ( all squarks are assumed degenerate
 in mass), and for various values of $\tan \, \beta$. (b) same as in (a)
 but for sleptons.\\[.15in]
Fig.5 : Total one--loop contribution (in percent), as a function of
$\tan \, \beta$, with $m_t=165$ GeV $m_{\tilde{q}}=400$ GeV and
$m_{\tilde{l}}=300$ GeV and three values of $\mu$.\\[.15in]
Fig.6 :One--loop sleptons contribution, as a function of
$\mu$ with $m_{\tilde{l}}= 300$ GeV, $m_{H^\pm}= 220$ GeV
and various values of $\tan \, \beta$.
\\[.15in]
Fig.7 : One--loop squarks contribution assuming the
scalar partners of the top quark to be non--degenerate,
 as a function of the mass of the heavier stop,
 with $\mu=250$ GeV, $m_{H^\pm}=220$ GeV, $m_t= 165$ GeV
and a left--right mixing angle $\theta_t = 0.4$ rad.,
(a) $\Delta m_{\tilde{t}} \equiv m_{\tilde{t}_1} -
m_{\tilde{t}_2} = 100$ GeV and $m_{\tilde{q}}= m_{\tilde{t}_1} - 50 GeV$
for the other squarks; (b) $\Delta m_{\tilde{t}}=300$ GeV and the other
squarks as in (a);\\[.15in]
Fig.8 : Feynman diagrams of the one--loop amplitude
with matter fields and their scalar
partners contributions. t, b (resp. $\tilde{t}, \tilde{b}$) denote
generically up and down quarks or leptons (resp. squarks or sleptons).
The nineth diagram is actually electron mass suppressed. The $W$
self-energies are needed for the $Z$ self-energie renormalization.
Tadpoles are fully subtracted through renormalization;\\[.15in]
Fig.9 : Contour plot for the total one--loop contribution to the
integrated cross section, in the $(\tan \, \beta, \mu)$ space and for
$m_t= 165$ GeV, $m_{\tilde{q}} = 400$ GeV $m_{\tilde{l}} = 300$ GeV
and $m_{H^\pm} = 220 $ GeV. All squarks (resp. sleptons) are assumed
degenerate in mass.
\end{document}